\documentclass[10pt,aps,twocolumn,superscriptaddress,longbibliography]{revtex4-1} 
\usepackage{latexsym}
\usepackage{graphicx}
\usepackage{amsmath}
\usepackage{amssymb}
\usepackage{verbatim}
\usepackage{braket}
\usepackage{upgreek}
\usepackage{float}
\usepackage{breqn}
\usepackage{subfig}
\usepackage{color}
\usepackage{hyperref}
\date{\today}
\makeatletter
\let\cat@comma@active\@empty
\makeatother
\begin{document}
	\title{Majorana Zero Modes and Bulk-Boundary Correspondence at Quantum Criticality}
	
	\author{S Rahul}
	\author{Ranjith R Kumar}
	\author{Y R Kartik}

	\affiliation{Department of Theoretical Sciences, Poornaprajna Institute of Scientific Research, 4, Sadashivanagar, 
Bangalore-560 080, India.}
	\affiliation{Graduate Studies, Manipal Academy of Higher Education, Madhava Nagar, 
Manipal-576104, India.}
	\author{Sujit Sarkar}
	\affiliation{Department of Theoretical Sciences, Poornaprajna Institute of Scientific Research, 4, Sadashivanagar, 
Bangalore-560 080, India.}
	
	\begin{abstract}
		\noindent 
			Majorana zero modes are well studied in the gapped phases of topological systems. We investigate Majorana zero modes at the topological quantum 
		criticality in one dimensional topological superconducting model with longer
		range interaction. We identify stable localized Majorana zero modes 
		appearing at criticality under certain conditions. Topological
		invariant number for these non-trivial 
		criticalities is obtained from zeros of a complex function
		associated with the Hamiltonian.
		Behavior of parametric curve at criticalities
		validate the invariant obtained and account for the appearance 
		of Majorana zero modes at criticality. Trivial and non-trivial topological
		nature of criticality due to the presence of multicritical point cause an unusual topological transition along the critical line. We observe and investigate this unique transition
		in terms of eigenvalue spectrum. Appearance of MZMs at criticality demands integer value of topological invariant number in order to validate the concept of bulk-boundary correspondence. Hence we propose a scheme to separate the invariant number into fractional and
		integer contribution to establish bulk-boundary correspondence at criticality.\\   


	\end{abstract}
	\maketitle

\section{Introduction}
In recent years, novel topological states of matter is
exponentially growing field in both theoretical and experimental condensed matter
physics \cite{stanescu2016introduction,wen2017colloquium,bansil2016colloquium,chen2019topological}.
This new phases of matter fall beyond the Landau's paradigm of spontaneous symmetry breaking
and order parameters.
In general, phase transitions are characterized by the emergence of singularities in the functions that constitute the physical parameters.
A quantum system possess an order parameter which dictates different phases and its divergence at the critical point signifies the quantum phase transition. These order parameters are system dependent i.e., different quantum systems possess different order parameters such as, magnetization for magnetic materials, density for liquid to gas transition etc \cite{landau}. In the plethora of quantum systems, topological quantum systems are a subclass which do not possess local order parameter, instead its characterization of topological phases are done in terms of a topological invariant number ($w$) which do not change under small perturbations until the energy gap closes. Change in the topological invariant number indicates the topological phase transition \cite{sachdev2007quantum}.\\
Among the vast variety of topological 
phases, one can identify an important class called symmetry protected 
topological phases
\cite{space,class,class1,hasan2010colloquium,pollmann2012symmetry}, 
where two quantum states have distinct topological properties protected 
by a same set of discrete symmetries \cite{altland1997nonstandard,S2019}. 
In these systems, the topological invariant number for the bulk Hamiltonian can be mapped to the number of edge states. This is 
know as the bulk-boundary correspondence \cite{CHEN2020126168,PhysRevE.100.020702,essin2011bulk}.
The topological invariant is quantized for gapped phases and it changes when there is a
topological quantum phase transition associated with the band gap closing.
For the topological systems such as short range Kitaev chain \cite{kitaev2001} and  SSH model \cite{su19795} (with only nearest neighbor coupling), there exists only a topological non-trivial ($w=1$) and trivial phases $(w=0)$. In this case the phase boundaries or criticalities does not display interesting features in terms of topology. However, when a system possesses more than one topological non-trivial phases \cite{niu2012majorana,sarkar2018quantization,kopp2005criticality}, one can investigate the criticality to understand the behavior of edge modes and find phenomenal features \cite{kumar2020multi,kartik2020topological}.\\ 
There are different edge modes with different physical properties that appear in the topological systems, like Massive Dirac edge modes \cite{massive}, Majorana zero modes \cite{kitaev2001,niu2012majorana,lutchyn2018majorana,kane} and Weyl fermions \cite{turner2013beyond}.
Among them, Majorana zero modes (MZMs) 
are localized at the edge of a one dimensional topological system when the bulk is gapped \cite{kitaev2001,wilczek2009majorana,nature2,cheng2011majorana,PhysRevB.92.075432}. However, at the criticality, the bulk gap itself is closed and the topological invariant number is ill-defined.
The ill-definition of topological invariant number hindered the study of MZMs and lack of physical 
understanding of bulk-boundary correspondence at criticality.\\
Nevertheless, topological invariant number have been defined at the 
criticality by omitting the transition 
point using infinitesimal parameter which results in 
fractional topological 
invariant number \cite{verresen2020topology}. 
There are several studies in the direction of understanding the criticality \cite{jones2019asymptotic,gapless,kestner2011prediction,
	keselman2015gapless,JIANG2018753,ganeshan2013topological,cano2015chirality,gao2019topological,wang2019non}. 
However, the question that still needs to be answered is about the validation of bulk-boundary correspondence
at criticality.\\
In this work, we intend to find the existence of localized 
stable MZMs living at the criticality and the corresponding 
characterization in terms of topological invariant number. We characterize all the critical lines using an alternate definition of invariant number and check the correspondence between the invariant number and the number of possible appearance of MZMs at criticality. 
The presence of MZMs at criticality indicates that it does not necessarily need bulk gap for their presence. 
Another motivation of this study is to understand the behavior of integral definition of topological invariant number and make an attempt to establish the definition at criticality. 
Although the literature suggests few studies on the criticality \cite{sarkar2018quantization,verresen2018topology,verresen2020topology}, we have made an attempt to understand the underlying 
structure of the integral form of the invariant number. This procedure shows that only integer part corresponds to the MZMs at criticality one has to discard the fractional part for its unphysicality with MZMs at 
criticality. 
By doing so we can retain only the integer part to validate bulk boundary correspondence since it has one to one correspondence with 
invariant number and number of MZMs at criticality.\\
This paper is organized as follows: In the Section \ref{model}, we 
introduce the model Hamiltonian and discuss critical lines
of topological trivial and non-trivial gapped phases and present the 
modified topological phase diagram with the characterization of both 
gapped and critical phases of the model.
We discuss the appearance of the MZMs at the criticality and
characterize all the critical lines using 
an alternate definition of topological invariant number in Section \ref{critMZMs}. 
We show the correspondence between the invariant number and the presence of MZMs at criticality. 
We also show that these MZMs at criticality are robust like the ones in gapped phases. We also
investigate a very interesting topological phase transition occurring 
along the critical line.
In Section \ref{BBC-PD}, we discuss the realization of 
bulk-boundary correspondence at criticality. Finally we summarize and conclude the results in Section \ref{conclude}.
\section{Model Hamiltonian}\label{model}
\noindent Kitaev chain with next nearest neighbor coupling is given by \cite{kopp2005criticality,niu2012majorana}, 
\begin{align} 
H = &-\mu \sum_{j=1}^{N} (1 - 2 c_{j}^{\dagger}c_{j}) - 
\lambda_1 \sum_{j=1}^{N-1} (c_{j}^{\dagger}c_{j+1} 
+ c_{j}^{\dagger}c_{j+1}^{\dagger} + h.c) \nonumber \\
&-\lambda_2 \sum_{j=2}^{N-1} ( c_{j-1}^{\dagger}c_{j+1} +  
c_{j+1} c_{j-1}+ h.c),
\label{jw} 
\end{align} 
where $c_{j}^{\dagger}$ and $c_{j}$ are fermionic creation and 
annihilation operators respectively.\@ $\lambda_1$, $\lambda_2$ and $\mu$ are 
nearest, next-nearest neighbor couplings and on-site chemical potential respectively.
Nearest-neighbor hopping amplitude $\lambda_1$ 
is also the amplitude of the nearest-neighbor superconducting gap, 
and the next-nearest-neighbor hopping amplitude $\lambda_2$ is equal 
to the next-nearest-neighbor superconducting gap.\\
Hamiltonian is diagonalized using Bogoliubov transformation to get Bloch Hamiltonian 
\cite{anderson1958coherent,sarkar2018quantization}, 
{\begin{equation}
	H_k = \boldsymbol{\chi} .\boldsymbol{\sigma}= \chi_{z} (k) \sigma_z - \chi_{y} (k) \sigma_y,
	\label{sudo}
	\end{equation} }
where $ \chi_{z} (k) = -2 \lambda_1 \cos k - 2 \lambda_2 \cos 2k + 2\mu,$ 
and $ \chi_{y} (k) = 2 \lambda_1 \sin k + 2 \lambda_2 \sin 2k.$\\
The excitation spectra can be obtained as 
\begin{equation}
E_k=\pm\sqrt{\chi_y(k)^2+\chi_z(k)^2}.
\end{equation}
Considering further longer range coupling,
gapped phases with higher invariant numbers can be obtained. 
Thus the model belongs to the topological class with invariant defined as  $\mathbb{Z}$, which can take integer values. Change in the value of $\mathbb{Z}$ (while preserving the 
symmetry) involves gap closing,
which are essentially the topological quantum phase transition points.
With nearest and next nearest neighbors in the model one can obtain three
such gap closing conditions, corresponding to which there exist three quantum critical lines.\\
The ground state 
energy is non-analytic at the excitation spectra $E_k=0$, signaling quantum phase transition. 
Three such points of this model in the
momentum space are at, $k=0$, $k=\pm\pi$ and $k=\cos^{-1}(-\frac{\lambda_1}{2\lambda_2})$.
Corresponding to these three points we have three 
critical lines $\lambda_2=\mu-\lambda_1$, $\lambda_2=\mu+\lambda_1$ and $\lambda_2=-\mu$ respectively 
(Refer to appendix.A for details). We study MZMs using model Hamiltonian in Bogoloubiv-de Gennes (BdG) form in open boundary setting which is given by, 
\begin{equation}
H_{BdG}  = \left(\begin{matrix}
\hat{h} & \hat{\Delta}\\
-\hat{\Delta} & - \hat{h}\\ 
\end{matrix}\right) 
\label{bdg}
\end{equation}
where submatrices are,
\begin{equation*}
\hat{h}_{ij} = \lambda_1 (\delta_{j,i+1} + \delta_{j,i-1}) + \lambda_2 (\delta_{j,i+2}+\delta_{j,i-2}) - 2 \mu \delta_{ij},\end{equation*}
\begin{equation}  \hat{\Delta}_{ij}=-\lambda_1 (\delta_{j,i+1} - \delta_{j,i-1}) - \lambda_2 (\delta_{j,i+2}-\delta_{j,i-2}).\label{B2}
\end{equation}
Eq.\ref{B2} are the components of the model Hamiltonian in the real basis, which yields 2N eigenvalues.
The spectrum of the BdG Hamiltonian always comes in pairs $\pm E$ because of the particle-hole symmetry. 
The form of particle-hole symmetry is, $P H_{BdG} P^{-1} = - H_{BdG}$ where the minus sign indicates the symmetric nature of the spectrum of the BdG Hamiltonian.\\
\begin{figure}[t]
	\centering
	\includegraphics[width=8.5cm,height=6.5cm]{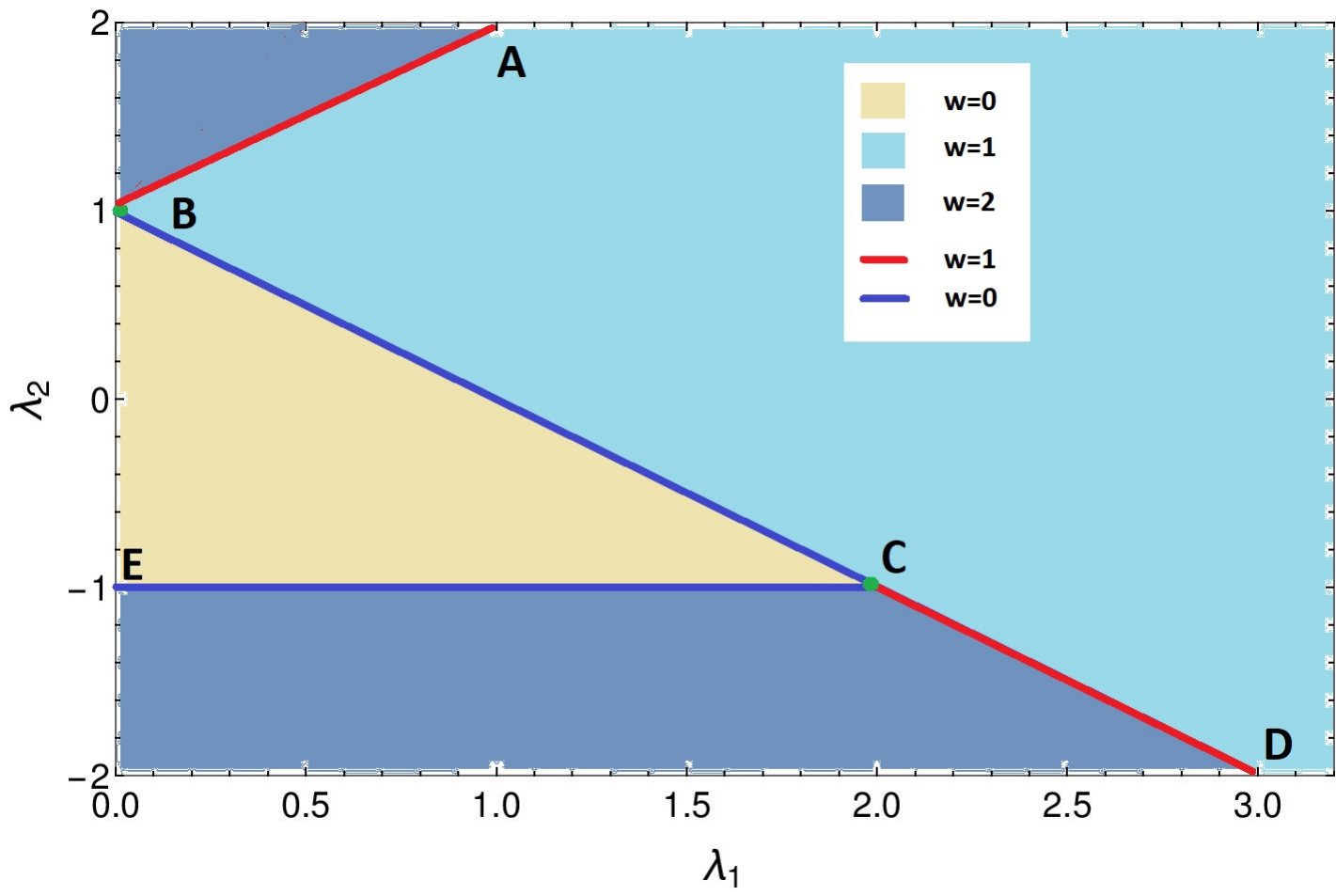}
	\caption{(Color online) Modified topological 
		phase diagram of the model Hamiltonian. Topological trivial and non-trivial
		criticalities are represented in blue and red lines respectively. Multicritical points
		are represented as green dots.}
	\label{Phasedia}
\end{figure} 
\noindent  Characterizing all gapped and gapless (critical) phases
of the model from the different analysis carried out in the further sections we 
represent the topological phase diagram Fig.\ref{Phasedia}. 
For the parameter regime, $\lambda_2>\mu+\lambda_1$ with $w=2$, 
$\lambda_2<\mu+\lambda_1$ and $\lambda_2>\mu-\lambda_1$ with 
$w=1$, $\lambda_2<\mu-\lambda_1$ and $\lambda_2 < -1$ 
with $w=0 $ and $\lambda_2<\mu-\lambda_1$ and $\lambda_2 > -1$ 
with $w=0 $, are the gapped phases of the system. The phase separations
$\lambda_2=\mu+\lambda_1$, $\lambda_2=\mu-\lambda_1$ and $\lambda_2=-\mu$
are gap closing critical points across which topological quantum
phase transition occurs.\\
The criticality $\lambda_2=\mu+\lambda_1$ (red line `AB') is characterized with
$w=1$, hence it is topologically non-trivial and posses localized MZMs. Similar 
characterization goes to the criticality $\lambda_2=\mu-\lambda_1$ for
$\lambda_1>2\mu$ (red line `CD'). On this critical line for $\lambda_1<2\mu$ (blue line `BC')
no MZMs are found and $w=0$ with the trivial characterization of criticality. 
The critical line $\lambda_2=-\mu$ (blue line `EC') also found to be trivial with $w=0$ since
no MZMs exist. The intersection point of two critical lines (green point `C') is
multicritical point, responsible for the topological transition between the 
distinct critical phases on the critical line $\lambda_2=\mu-\lambda_1$.\\
Authors of Ref. \cite{niu2012majorana} has explored the topological phase diagram of this model under both broken and unbroken 
time-reversal invariance. They have also explored the MZMs at the gapped phases and the topological invariant 
which characterize them. In our study we focus on the zero modes at the criticality which distinguish them into trivial and non-trivial topological nature. Authors of Ref. \cite{sarkar2018quantization} has also explored the criticality in terms of conventional 
topological invariant and reported the possibility of fractional invariants. However, the bulk-boundary correspondence 
has not been taken into consideration during this study thus it is not the complete physical picture. We 
overcome this problem and explore the possible characterization 
of the zero modes at criticality and the corresponding physical picture of bulk-boundary correspondence.
\section{Topological characterization of Majorana zero modes at criticality}\label{critMZMs}
\begin{figure}[t]
	\centering
	\includegraphics[scale=0.25]{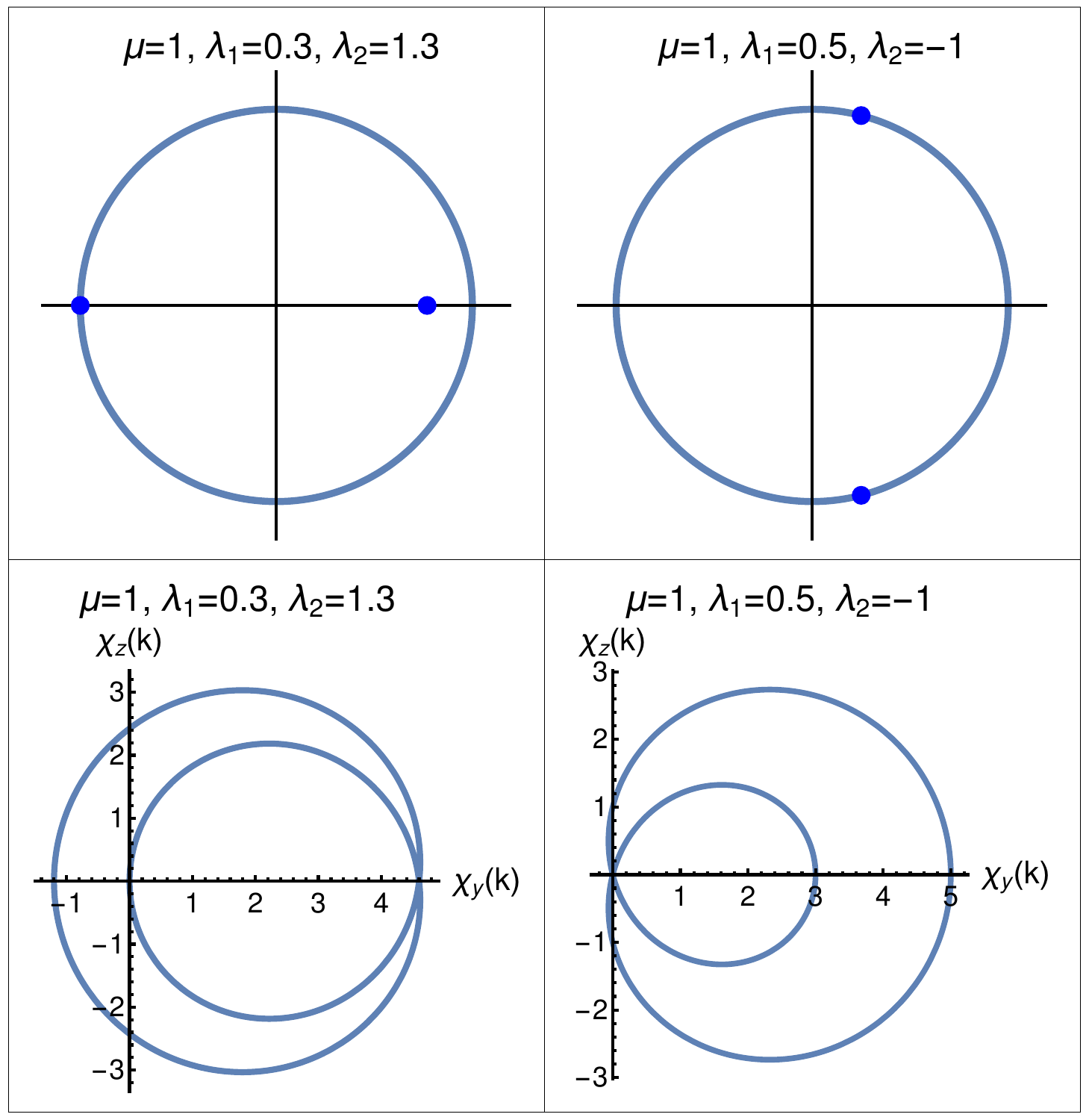}
	\caption{(Color online) Upper panel corresponds to zeros of complex function (Eq.\ref{sol}) 
		for the critical lines $\lambda_2 = \mu + \lambda_1$ (left figure) and $\lambda_2=-\mu$ (right figure). 
		Zeros inside the unit circle corresponds to existence of MZMs. 
		Lower panel corresponds to parametric curves plotted at critical lines $\lambda_2 = \mu + \lambda_1$ (left figure) and $\lambda_2=-\mu$ (right figure).}
	\label{unit000}
\end{figure}

\begin{figure}[t]
	\centering
	\includegraphics[width=8cm,height=6cm]{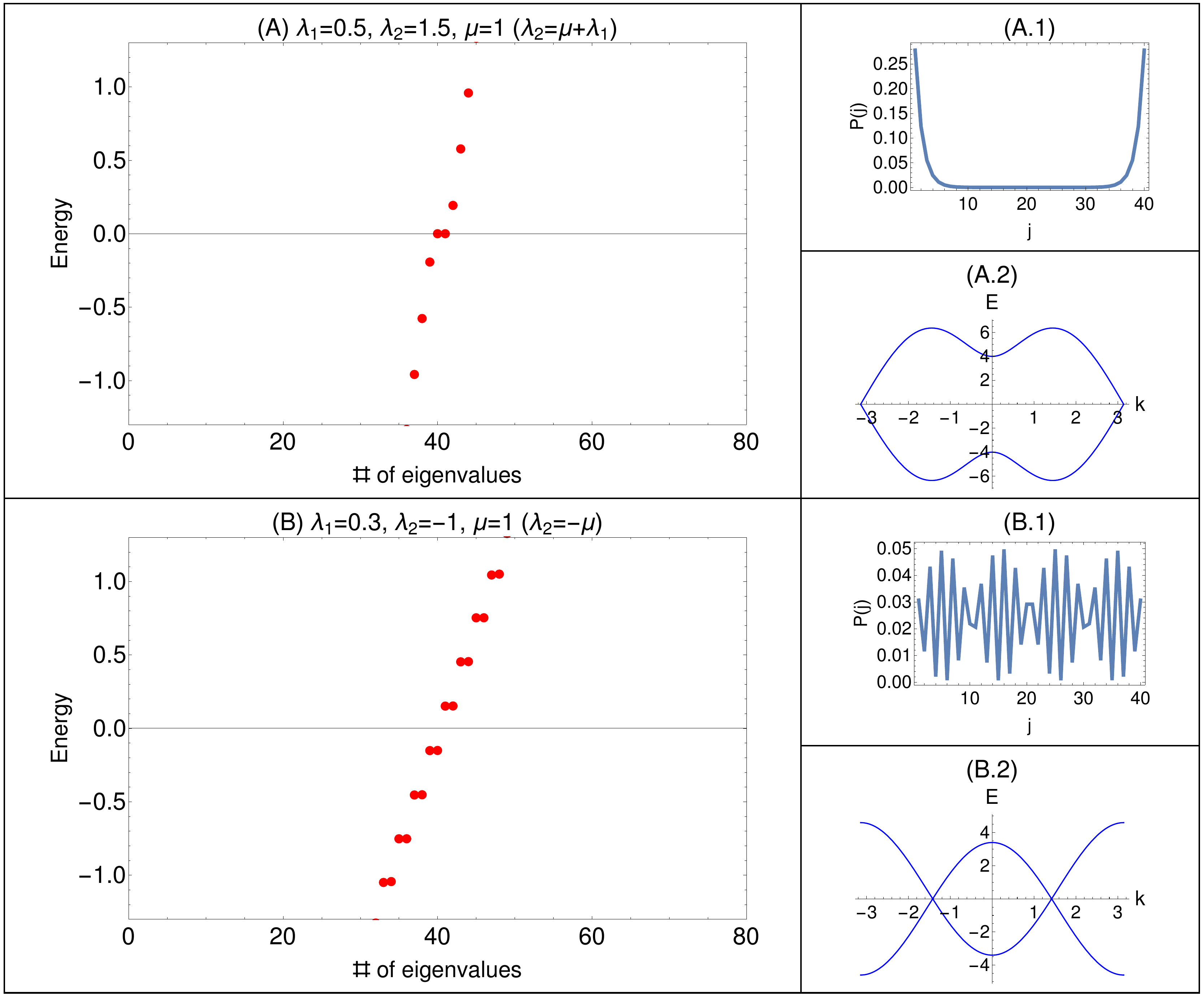}
	\caption{(Color online) Upper and lower figures corresponds to eigenvalue distribution (A, B), probability distribution (A.1, B.1) and 
		excitation dispersion (A.2, B.2) at criticalities $\lambda_2=\mu+\lambda_1$ and $\lambda_2=-\mu$. Numerical diagonalization was carried out for system size $N=40$ in open boundary setting. The $2N$ eigenvalues are obtained from the real space representation of BdG Hamiltonian Eq.\ref{B2} but in the plots A and B, we focus on the zero modes which lie on zero energy axis and hence we restrict the spectrum to $-1.5$ to $1.5$.}
	\label{Pr6}
\end{figure}

\begin{figure}[t]
	\centering
	\includegraphics[width=8cm,height=5cm]{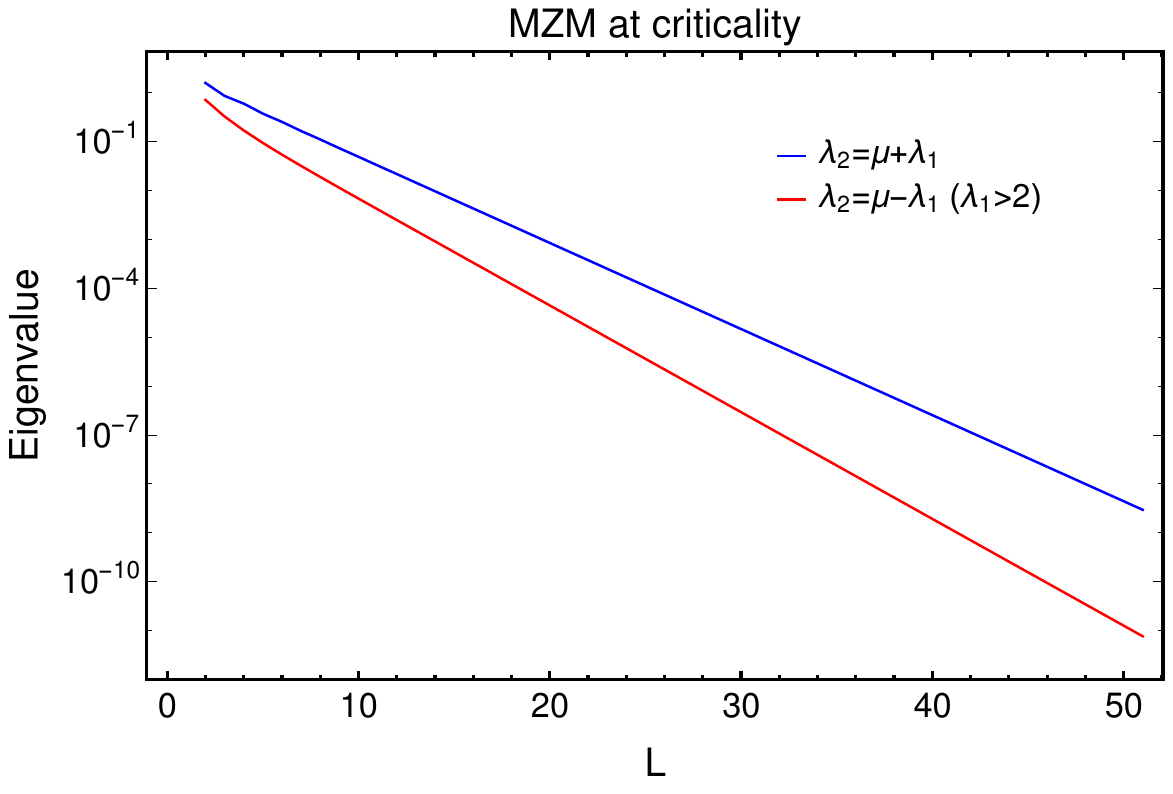}
	\caption{(Color online) System size dependence of Majorana zero modes at criticality. Eigenvalues in the y axis are plotted in the log scale. } 
	\label{stab}
\end{figure}

\begin{figure}[t]
	\centering
	\includegraphics[width=8cm,height=5cm]{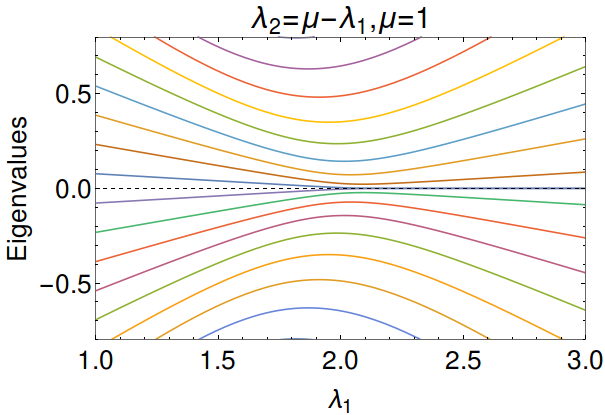}
	\caption{(Color online) Eigenvalue spectrum of the Hamiltonian as a function of coupling 
		parameter $\lambda_1$ on a critical line $\lambda_2 = \mu-\lambda_1$. 
		Dashed line is the axis of zero modes (zero energy eigenvalues). Spectrum is plotted for the system size $N=40$ in the open boundary setting.}
	\label{ext3}
\end{figure}

\noindent Critical lines are the phase boundaries that separates topological 
gapped phases. The edge mode physics at the gapped phases can be characterized 
by calculating topological invariant number, which is the essence of 
bulk-boundary correspondence. This conventional 
topological invariant number is ill-defined, because the bulk gap closes 
at criticality. 
Hence, in this work we make use of an alternate definition of the 
topological invariant number, using zeros and poles of a complex function 
associated to the model Hamiltonian, to characterize all 
the critical lines. \\
The model Hamiltonian in Eq.\ref{jw} can be 
written in the Majorana basis, using $c_j^{\dagger}=\frac{a_{j}+ib_{j}}{2},c_j=\frac{a_{j}-ib_{j}}{2}$, as  
\begin{equation}
H = -i\sum_{\alpha=0,1,2} \left( \sum_{j=1}^{N-\alpha}\gamma_{\alpha} b_j a_{j+\alpha}\right) ,\label{MJ2}
\end{equation}
where $a_j$ and $b_j$ are Majorana operators satisfying anti-commutation 
relation, $\gamma_{0}=-\mu$, 
$\gamma_{1}=\lambda_1$ and $\gamma_{2}=\lambda_2$.
The Hamiltonian can be translated into the Fourier space as 
$f(k)= \sum_{\alpha=0,1,2} \gamma_{\alpha} e^{ik\alpha}$.
Considering
$z = e^{ik}$ and interpreting $f(k)$ on the unit circle in 
a complex plane, the complex function 
associated with the Hamiltonian can be written as (we refer to appendix.C for detailed derivation)
\begin{align} 
f(z) = \sum_{\alpha=0,1,2} \gamma_{\alpha} z^{\alpha} =-\mu + \lambda_1 z +\lambda_2 z^2.
\label{so}
\end{align}   
The solutions of this function $z_1$ and $z_2$ are
\begin{equation}
z_{1,2}= \frac{-\lambda_1 \pm \sqrt{\lambda_1^2+4\mu \lambda_2}}{2\lambda_2},
\label{sol}
\end{equation}
which are the two zeros of the function $f(z)$. 
Topological invariant ($w$) is defined to be equal to the number 
of zeros ($N_z$) and order of the pole ($N_p$) inside the unit circle \cite{ablowitz2003complex}. 
It can be written as
\begin{equation}
w=N_z-N_p.
\end{equation}  
In this case the winding number equals to the zeros inside unit circle
since there is poles in Eq.\ref{so} (i.e, $N_p=0$).
Zeros inside the circle represents 
topological phase, on the circle represents transition point or criticality, 
and outside the circle represent non-topological phase. 
The winding number calculated in this way, characterize the edge modes
both in gapped phases as well as on the critical lines.\\
There are three distinct gapped phases $w=0,1$ and $2$ in our model.
Correspondingly, one can find the zeros $z_1$ and $z_2$
inside or outside the unit circle. For the gapped phase $w=0$, both
$z_1$ and $z_2$ lie outside the unit circle representing 
topological trivial phase. For the gapped phase $w=1$, one of
the zero $z_1$ lie inside and the $z_2$ lie outside representing
topological non-trivial phase with one MZMs at each end of the chain.
Similarly for the gapped phase $w=2$ one has both the zeros lie inside 
the unit circle representing two MZMs at each end of the chain. This 
observation of the zeros of the complex function associated to the 
Hamiltonian validate the method in characterizing the winding number and 
to determine the topological triviality and non-trivialities. 
Based on this observation we now apply the method to characterize the
criticalities.\\
Our model Hamiltonian consists of three 
critical lines $\lambda_2=\mu-\lambda_1$, $\lambda_2=\mu+\lambda_1$  
and $\lambda_2=-\mu$ which separates $w=0,1,2$ topological gapped phases.
At first, we analyze the results for the critical lines $\lambda_2=\mu+\lambda_1$
and $\lambda_2=-\mu$ which is shown in Fig.\ref{unit000}. It consists of two panels.
Upper panel shows the behavior of winding
number in terms of zeros
of the complex function around the unit circle at criticality. 
Lower panel shows the parameter space representation of Anderson pseudo spin Hamiltonian
at criticality. There are two loops in the parametric curve since in both transitions
$w=2$ phase is involved i.e,
the transitions are between $w=0$ or $1$ and $w=2$. 
In both the panels left and right figures corresponds to
$\lambda_2=\mu+\lambda_1$ and $\lambda_2=-\mu$ critical lines respectively.
Notice that, in the upper panel, one of the two zeros (blue dots) is on 
the unit circle indicating that the system is at criticality. Similarly in the 
lower panel, one among the two loops of the parametric curve intersecting 
the origin signals the criticality. 
It is evident that for $\lambda_2=\mu+\lambda_1$ there is non-trivial winding number
($w=1$) since one of the zero lies inside the unit circle. Thus there exists a localized
MZMs at the criticality $\lambda_2=\mu+\lambda_1$. The analysis of its stability in 
comparison with gapped phases is discussed later. In the parametric curve,
the non-trivial winding number at criticality is mimicked as the winding of 
one of the loop enclosing the origin, while another intersecting the origin.
In the spirit of the principle of bulk-boundary correspondence this implies 
a novel phenomenon of emergence of localized edge modes at criticality,
which exists even without the bulk gap. Hence the critical line $\lambda_2=\mu+\lambda_1$
can be regarded as topologically non-trivial criticality. 
Emergence of these unconventional edge modes are not possible at all
the critical lines. Some criticality remain topologically trivial with
winding number $w=0$. In our model the critical line $\lambda_2=-\mu$
remain trivial as shown in the left figures in the upper and lower panels
of Fig.\ref{unit000}. In this case both zeros lie on the unit circle representing
the trivial criticality with $w=0$, in other words with no localized edge modes.
This can be verified from the parametric curve, where both the loops intersect the origin
signaling the trivial criticality with zero winding. Unlike in other critical
lines, here the zeros are imaginary numbers.\\
In support to the previous characterization, presence of 
MZMs at criticality can be verified numerically in
open boundary condition. We investigate two critical lines in terms of eigenvalues, probability distribution of wavefunction along with gap closing excitation dispersion and look for MZMs. 
The upper and lower panels, (A) and (B) of the Fig.\ref{Pr6}, depicts the presence and absence of MZMs at critical lines
$\lambda_2=\mu+\lambda_1$ and $\lambda_2=-\mu$ respectively 
for open boundary condition.
One can observe the presence of a pair of eigenvalues (red dots)
at zero energy for $\lambda_2=\mu+\lambda_1$ as shown in Fig.\ref{Pr6}(A).  Finite probability distribution of wavefunction at the edge (see  Fig.\ref{Pr6}(A.1)) is observed which represents the presence of MZMs at the critical line. In support to the pair of eigenvalues representing MZMs at criticality, the gap closing at $k_0=\pm\pi$ in the momentum space is also shown Fig.\ref{Pr6}(A.2).
These zero modes are absent in the case of $\lambda_2=-\mu$ (see Fig.\ref{Pr6}(B)), where there is no
zero energy eigenvalues on the zero energy line. The corresponding probability distribution of the MZMs is not finite at the edge as shown in Fig.\ref{Pr6}(B.1) and also the criticality is associated with the gap closing at $k_0=\pm 1.42$ (see Fig.\ref{Pr6}(B.2)) which corresponds to absence of MZMs.\\
\noindent Topological non-triviality of certain critical lines and triviality of others can 
be understood by analyzing the delocalization of MZMs into the bulk 
as the topological transition occurs between gapped phases. It is well 
understood that as one approaches the criticality, the localization length of MZMs 
diverges, as a consequence the zero mode merge into the bulk. This notion is true 
if the system is undergoing the transition between topologically trivial $(w=0)$ and non-trivial $(w=1)$
gapped phases. When there is a transition between two non-trivial gapped phases
with distinct winding number, we observe that MZMs remain localized
even at the criticality. The number of zero modes localized at criticality
(i.e, the winding number) is equal to the lowest winding number among 
the two non-trivial gapped phases between which the 
transition is occurring. In our model, for the transition between two
non-trivial gapped topological phases $w=1$ and $w=2$, one of 
the MZMs remain localized at the criticality $\lambda_2=\mu+\lambda_1$,
which is the line of separation between these phases.
The triviality of the critical line $\lambda_2=-\mu$ can now be 
understood easily as the least invariant number between $w=0$ and $w=2$ is $w=0$. 
Both the MZMs are forced to merge with the bulk as one approaches the criticality.\\
Meanwhile, critical line $\lambda_2=\mu-\lambda_1$ shows a very unique feature.
The critical line separates, the gapped phases $w=0$ and $w=1$
for $\lambda_1<2\mu$ and the gapped phases $w=2$ and $w=1$ for $\lambda_1>2\mu$.
This is the consequence of the intersection of critical line $\lambda_2=-\mu$
with $\lambda_2=\mu-\lambda_1$ at $\lambda_1=2\mu$.
This intersecting 
point is a multicritical point at which all three gapped phase
$w=0,1$ and $2$ meet. On the critical line $\lambda_2=\mu-\lambda_1$
the multicritical point separates the trivial
and non-trivial criticalities. Thus a single critical line has both 
topological and non-topological nature. This intriguing
feature result in the unusual topological transition between distinct criticalities.
We first obtain the winding number and MZMs behavior at this criticality as
obtained previously for other critical lines.\\
\begin{figure}[t]
	\centering
	\includegraphics[width=7.5cm,height=5cm]{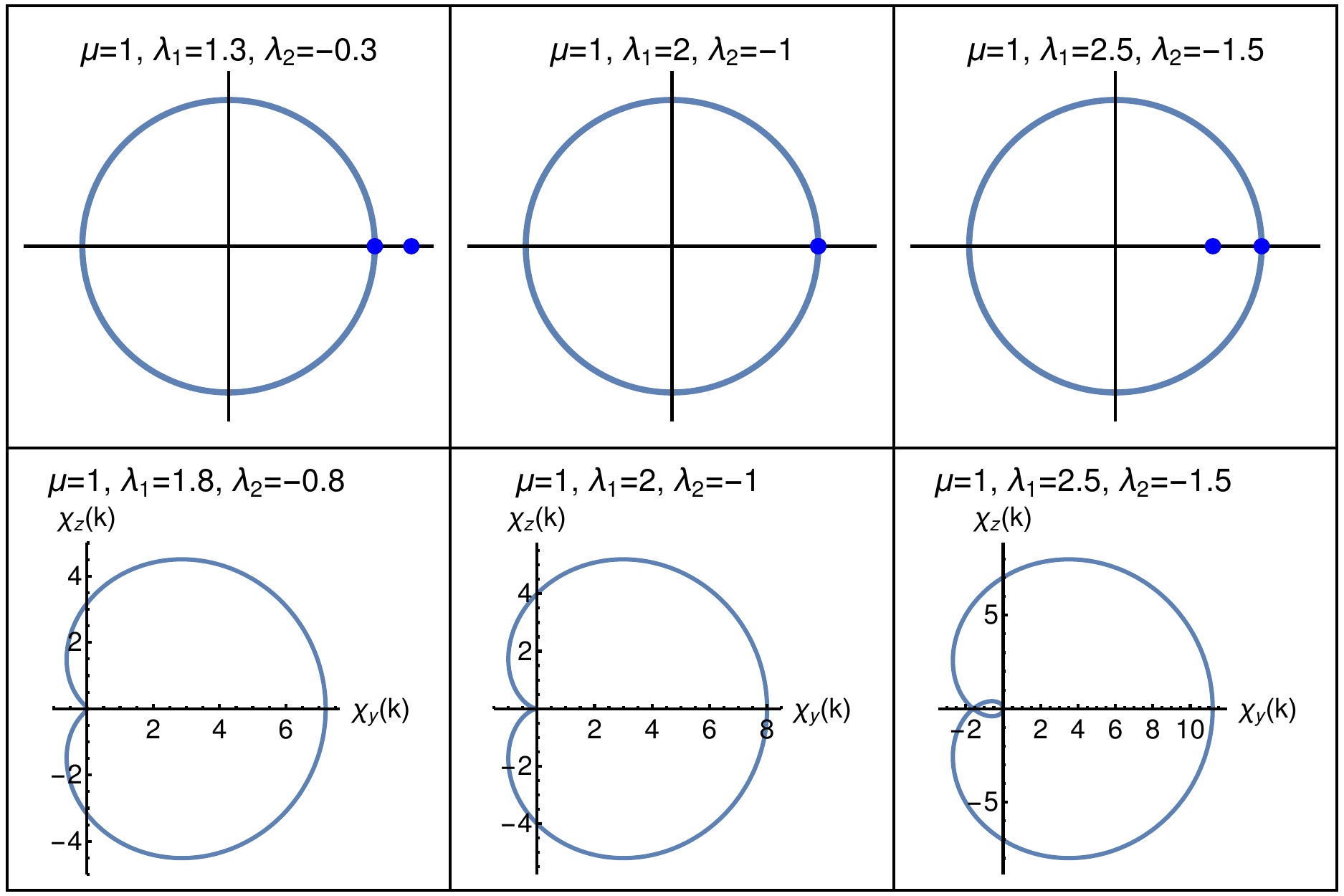}
	\caption{(Color online) Upper panel corresponds to zeros of complex function (Eq.\ref{sol}) 
		for the critical line $\lambda_2 = \mu - \lambda_1$ (left, middle and right figure). 
		Zeros inside the unit circle corresponds to existence of MZMs. 
		Lower panel corresponds to parametric curves plotted at 
		critical line $\lambda_2 = \mu - \lambda_1$ (left, middle and right figure).
		Left, middle and right plots of both panels
		are respectively for $\lambda_1<2\mu$ (trivial), 
		$\lambda_1=2\mu$ (multicritical point) and $\lambda_1>2\mu$ (non-trivial) 
		regimes on the critical line $\lambda_2=\mu-\lambda_1$.}
	\label{TQC}
\end{figure}
On the critical line $\lambda_2=\mu-\lambda_1$, winding number 
is $w=0$ for $\lambda_1<2\mu$ region 
and is $w=1$ for $\lambda_1>2\mu$ region. This can be 
inferred from the upper panel of the Fig.\ref{TQC}, which shows 
the behavior of zeros of the complex function associated to the Hamiltonian.
For $\lambda_1<2\mu$, one of the zero lies on the 
unit circle representing the criticality, while other zero lies outside
the unit circle representing trivial winding number $w=0$ as shown in
the left plot of Fig.\ref{TQC}. The middle plot is for $\lambda_1=2\mu$, 
represent multicritical point with both the zeros
lie at the same point on the unit circle. Topological non-trivial
criticality is obtained for  $\lambda_1>2\mu$, as
shown in right plot of Fig.\ref{TQC}, where one of the zero lies
inside the unit circle implying non-trivial winding number $w=1$.
Thus both topological trivial and non-trivial criticalities separated 
by a multicritical point is obtained on the critical line $\lambda_2=\mu-\lambda_1$.
Parametric curves in the lower panel supports this results, where for 
trivial criticality single loop intersecting the origin, for 
multicritical point the loop gets sharper near the intersection
to the origin, for non-trivial criticality there are two loops and
one among them showing winding enclosing the origin, is obtained (see respectively left, middle and right plots in the lower panel of Fig.\ref{TQC}).\\
\begin{figure}[t]
	\centering
	\includegraphics[width=8cm,height=6cm]{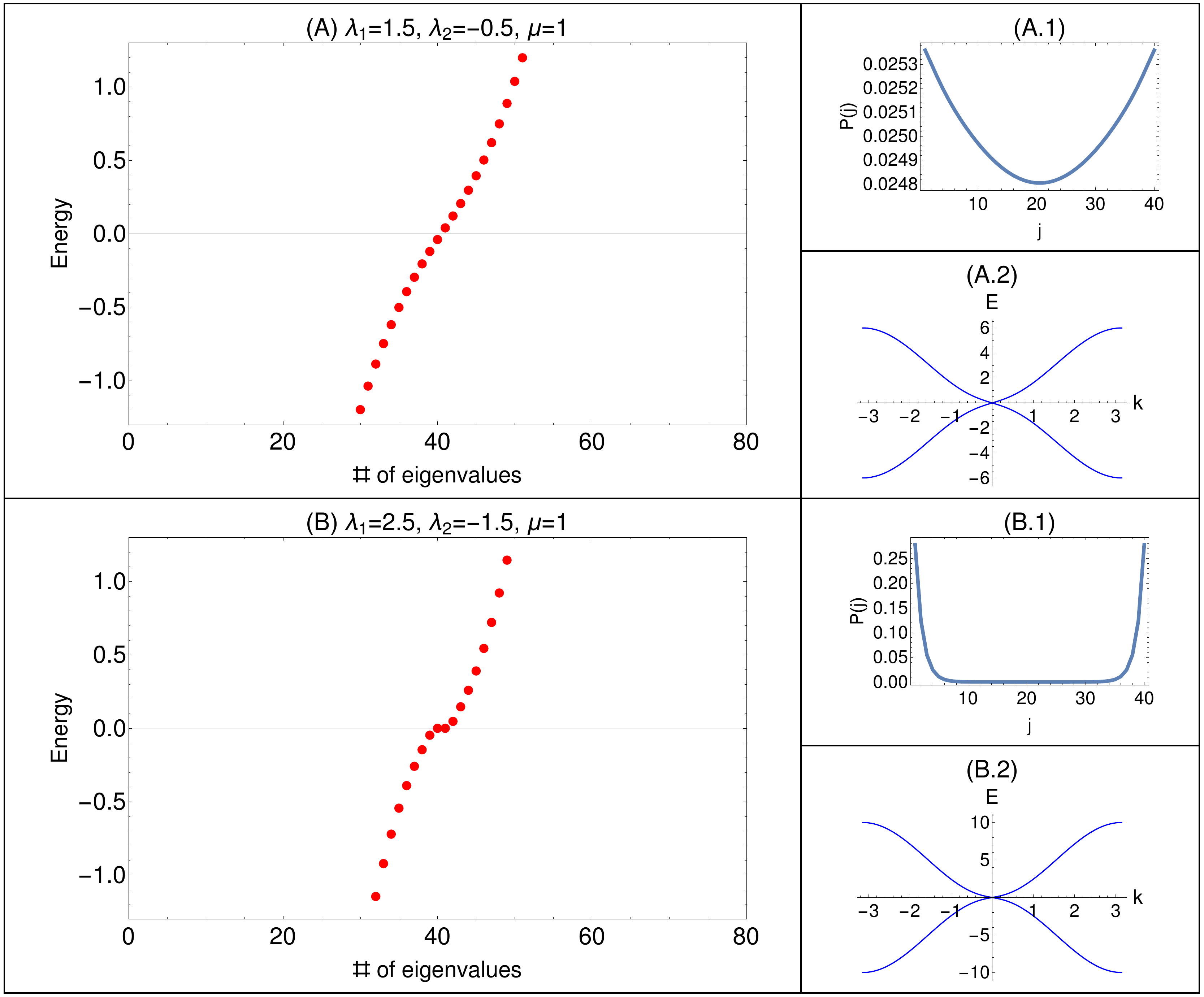}
	\caption{(Color online) Upper and lower figures corresponds to eigenvalue distribution (A, B), probability distribution (A.1, B.1) and 
		excitation dispersion (A.2, B.2) at criticality $\lambda_2=\mu-\lambda_1$. Specifically 
		Upper and lower figures are respectively for
		$\lambda_1<2\mu$ and $\lambda_1>2\mu$. Numerical diagonalization was carried out for system size $N=40$ in open boundary setting. The $2N$ eigenvalues are obtained from the real space representation of BdG Hamiltonian Eq.\ref{B2} but in the plots A and B, we focus on the zero modes which lie on zero energy axis and hence we restrict the spectrum to $-1.5$ to $1.5$.}
	\label{ext1}
\end{figure}   
MZMs can be observed in the Fig.\ref{ext1}(A) for 
same parameters space as obtained analytically. Fig.\ref{ext1}(A)
shows the localized MZMs for $\lambda_1>2\mu$ on the critical line $\lambda_2=\mu-\lambda_1$.
Pair of zero energy eigenvalues and the corresponding finite of probability
distribution (see Fig.\ref{ext1}.(A.1))
of MZMs are the convincing indicators of topological non-trivial criticality. Excitation dispersion with gap closing, in Fig.\ref{ext1}(A.2), is presented to show the parameter space considered here is on the critical line associated with gap closing at $k_0=0$.
On the other hand, MZMs are absent for $\lambda_1<2\mu$ on the same critical line as
shown in Fig.\ref{ext1}(B). There are no zero energy eigenvalues as well as finite probability distribution 
at the edge (Fig.\ref{ext1}(B.1)). Nevertheless the criticality remains same with gap closing at momentum $k_0=0$ (Fig.\ref{ext1}(B.2)). This represents the trivial
criticality with no localized MZMs. The peculiar behavior of 
the MZMs on this critical line can also be understood from the 
nature of gapped phases that reside in the region $\lambda_1>2\mu$ and
$\lambda_1<2\mu$.\\
The presence of MZMs at criticality raises the question on the fragility of these MZMs at the topological quantum critical lines? 
At the topological phase transition point, since the bulk gap is closed, eigenvalues tend to distribute equally across zero energy axis. In this case, there are possibilities of eigenvalues lying close to zero energy axis mimicking the nature of MZMs. By doing the system size dependence study of such eigenvalues reveal whether they are truly lying on the zero energy axis or just another bulk eigenvalue.\\ 
In Fig.\ref{stab}, we study system size dependence of MZMs eigenvalue to show 
its robustness at criticality. 
\noindent  
From Fig.\ref{stab}, it is evident that zero mode eigenvalues plotted with respect to system size at criticality shows the negative slope. For an increase in the system size, the zero mode eigenvalues at different criticalities show the exponential decay which is the characteristic feature of MZMs in general.\\ 
Topologically trivial and non-trivial characters of criticality
on a single critical line $\lambda_2=\mu-\lambda_1$ give rise to an unique
\textit{topological transition at quantum criticality}.
This topological phase transition occur through a multicritical 
point independent 
of the bulk gap closing. This intriguing feature is depicted in Fig.\ref{ext3}.
\noindent It shows the eigenvalue 
spectrum with respect to the coupling parameter $\lambda_1$ along the critical line
$\lambda_2 = \mu - \lambda_1$. Bulk remain gapless representing
the criticality. Setting $\mu=1$, we observe no zero modes in the region $0<\lambda_1<2$. 
Zero mode appear for $\lambda_1>2$, which imply that the
there exists distinct gapless (critical) phases and the transition occur among them 
through multicritical point at $\lambda_1=2$ acting as the transition point. This transition is topological
quantum phase transition since the distinct critical phases can be characterized 
as topological ($w=1$) and non-topological ($w=0$) in nature as discussed earlier.\\ 
\noindent The zero mode solutions of the Hamiltonian in Eq.5, for both gapped phases and at criticality can be obtained from Eq.\ref{9} given as, \begin{equation*}
e^{q} = \frac{-\lambda_1 \pm \sqrt{\lambda_1^2 + 4 \lambda_2 \mu}}{2 \lambda_2} 
\end{equation*}
For the non-topological phase, the roots satisfy the condition, $e_{1}^{q} >1 $, $e_{2}^{q} >1$ and for the topological phase, the roots satisfy the conditions
(i) $e_{1}^{q} <1 $, $e_{2}^{q} <1$ (for $w=2$),
(ii) $e_{1}^{q} <1 $, $e_{2}^{q} >1$ or vice versa (for $w=1$).
\begin{table}[ht]\caption{Analytical results at criticality}
	\centering 
	\begin{tabular}{c | c c c}
		\hline\hline                      
		& & No. of MZMs & Condition \\ [0.5ex]
		\hline                 
		Gapped	&w=0 & 0 & $e_{1}^{q} >1 $, $e_{2}^{q} >1$ \\
		phases&w=1 & 1 & $e_{1}^{q} <1 $, $e_{2}^{q} >1$  \\
		&w=2 & 2 & $e_{1}^{q} <1 $, $e_{2}^{q} <1$ \\ 		
		\hline\hline
		&$\lambda_2 = \mu + \lambda_1$ & 1 & $e_{1}^{q} <1 $, $e_{2}^{q} >1$\\
		Criticality	&$\lambda_2 = -\mu$ & 0 & $e_{1}^{q} >1 $, $e_{2}^{q} >1$\\
		&$\lambda_2 = \mu - \lambda_1$ $(\lambda_1 <2)$ & 0 & $e_{1}^{q} >1 $, $e_{2}^{q} >1$\\
		&$\lambda_2 = \mu - \lambda_1$ $(\lambda_1 >2)$ & 1 & $e_{1}^{q} <1 $, $e_{2}^{q} >1$\\[1ex]
		\hline\hline
	\end{tabular}\label{table:nonlin}
\end{table}
The presence of MZMs at both gapped phases and criticality can be verified analytically from the Eq.\ref{9} which is shown in the Table.\ref{table:nonlin}. These roots corresponds to MZMs at the end of the chain. For $w=1$ phase, one of the root will be less than 1 and other greater. With this notion, the presence of MZMs at criticality can also be verified. Analytical results show that at criticalities $\lambda_2 = \mu + \lambda_1$ and $\lambda_2= \mu- \lambda_1$ $(\lambda_1 >2)$ host MZMs and criticalities $\lambda_2 = -\mu$ and $\lambda_2 = \mu - \lambda_1$ $(\lambda_1 <2)$ does not host MZMs. These results can also be verified from our previous discussion on numerical analysis of MZMs at criticality.
\section{Bulk boundary correspondence}\label{BBC-PD}
\begin{figure}[t]
	\centering
	\includegraphics[width=8cm,height=3.5cm]{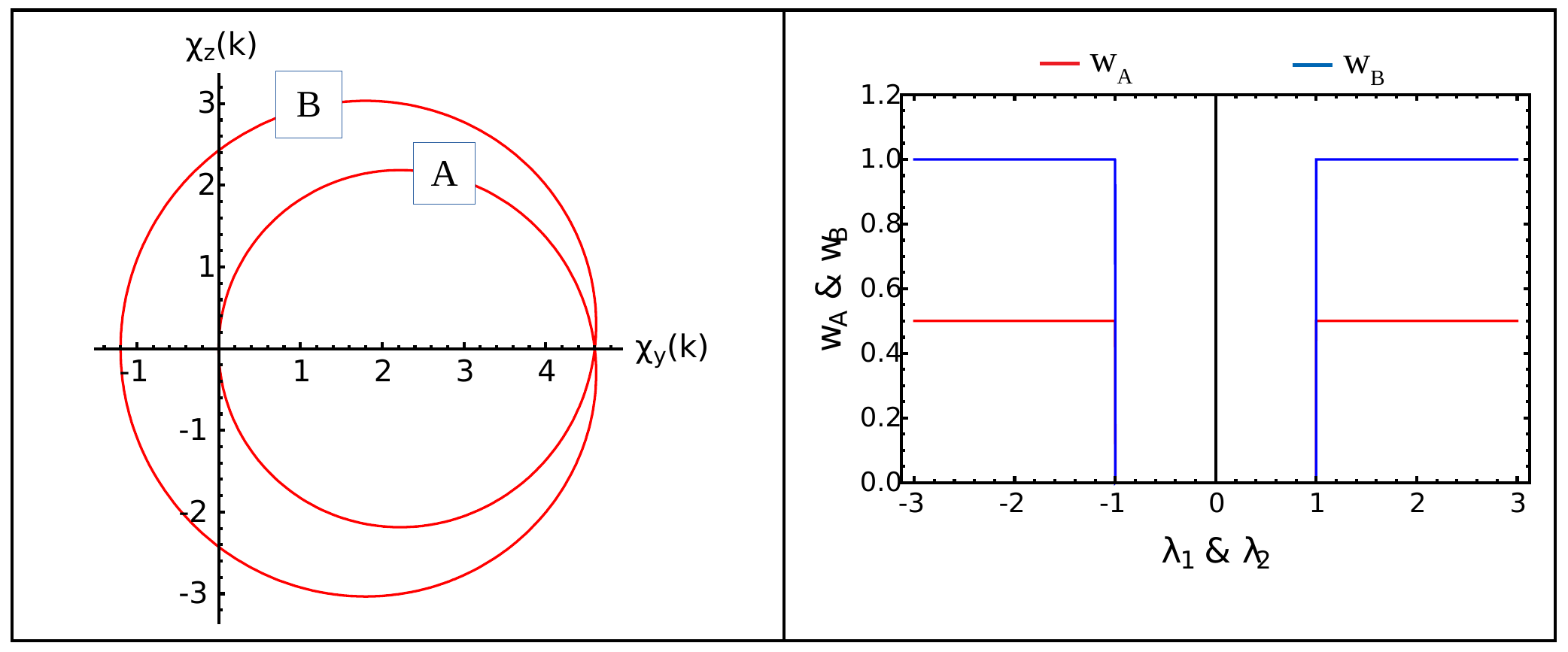}
	\caption{(Color online) Left and right panel corresponds to generic parametric curve representation at non-trivial criticality and fractional ($w_A$) and integer ($w_B$) contributions of the topological invariant number respectively.}
	\label{dcir}
\end{figure}

\noindent In this section we study the shortcomings of the integral definition of topological invariant number at criticality. We investigate the underlying structure of the integral definition and necessary modifications for bulk boundary correspondence to be valid at criticality. In general,
the definition of the topological invariant number can be written as 
\begin{equation}
w = \frac{1}{2\pi}\oint A_k dk, \label{wind}
\end{equation}
where $A_k$ is the Berry connection \cite{10.2307/2397741}. Berry connection for a given Bloch wavefunction $\psi_k (r) = u_k (r) e^{ikr}$ is defined as, 
$A_k = i <u_k|\partial_k|u_k>$ and $w$ is winding 
number in this case.
This winding number is valid only in gapped phases where it takes 
integer values. 
At the criticality the winding number in Eq.\ref{wind} is ill-defined since the Berry connection $A_k$ is indeterminate.
In spite of this, the integral has been computed which resulted in fractional invariant numbers, as in the Ref.\cite{sarkar2018quantization}. The author has 
obtained fractional invariant number at criticality but does not take into account 
the singularity during the integration and also does not address anything on the bulk-boundary correspondence at criticality.
One can also obtain $w$ through the integration by omitting the singularity, which will again
results in the fractional values at criticality \cite{verresen2020topology}. 
In the Ref.\cite{verresen2020topology}, the author has obtained the 
fractional invariant number at criticality and also it addresses the issue of singularity in the integral.
However, the fractionality of the topological
invariant number does not corresponds to the number of MZMs
at the edge thus the study lacks in establishing one to one correspondence between MZMs present in the system and topological invariant number at criticality.
Hence our motivation is to understand the underlying structure of the integral form of the 
topological invariant number at criticality.
Therefore, we propose a method where one has to separate out the integer and fractional parts from the integral form to get a one to one correspondence between topological invariant number and number of MZMs at criticality.  \\
Topological invariant number at a criticality can be obtained by omitting an infinitesimal neighborhood of a gap closing point \cite{verresen2020topology}, reads
\begin{equation}
w=\frac{1}{2\pi} \lim_{\epsilon\rightarrow 0} \int_{|\chi|>\epsilon} \mathcal{A}_k \;\; dk,
\label{fracwind}
\end{equation}
where $\boldsymbol{\chi}$ is Anderson pseudo spin vector and $\mathcal{A}_k= \frac{\chi_{z} \partial_k \chi_{y} - \chi_{y} \partial_k \chi_{z}}{\chi_{z}^2+\chi_{y}2}$ is
Berry connection in our case. The winding number $w$ acquires 
fractional values at criticality in contact to the integers in the 
gapped phases. Fractional invariants numbers does not corresponds
to physical edge modes in the system. To overcome this we propose to separate 
the fractional and integer contribution of the Berry connection in the 
integral.\\
A generic parametric curve representing non-trivial criticality is shown in left panel of 
Fig.\ref{dcir}. There exists localized zero mode at this criticality correspondingly
the non-trivial winding $w=1$. Analyzing the parametric curve one can easily 
spot two loops (named as `A' and `B'), one intersecting the origin and another 
enclosing the origin. 
It is evident that the loop intersecting the origin contributes the fractional
part and the loop enclosing the origin contributes the integer part in the 
invariant number. Two loops in our case originates from the first and second 
nearest neighbor interactions. It is reasonable to split $\mathcal{A}_k$
as the sum of the contributions from first and second nearest neighbors.
It can be written as
\begin{equation}
\mathcal{A}_k= \mathcal{A}_k^{NN} + \mathcal{A}_k^{NNN},
\end{equation}
where $\mathcal{A}_k^{NN}$ and $\mathcal{A}_k^{NNN}$ are for first and second nearest neighbors
respectively.
In our model non-trivial criticalities $\lambda_2=\mu+\lambda_1$
and $\lambda_2=\mu-\lambda_1$ (for $\lambda_1>2\mu$) and the 
localized zero modes can be efficiently characterized in this method.
During the transition between gapped phases $w=2$ and $1$, 
invariant number at criticality is obtained as $w=\frac{3}{2}$ from
Eq.\ref{fracwind}. Using the method of separation of nearest
neighbor contributions in the integral we obtained the
fractional invariant number as the sum of $w=1$ ($w_B$) and $w=\frac{1}{2}$ ($w_A$)
as shown in the right panel of Fig.\ref{dcir}. \\
One can write $\mathcal{A}_k$, i.e., \begin{equation}\mathcal{A}_k = \frac{\lambda_1^2 + 2 \lambda_2^2 + \lambda_1 (3 \lambda_2 - \mu) \cos k - 2 \lambda_2 \mu \cos 2k}{2 \left(\lambda _1^2+\lambda_2^2+\mu ^2 + 2 \lambda_1 \cos k (\lambda_2-\mu )-2 \lambda_2 \mu  \cos 2 k\right)}\end{equation} in a separate manner dividing contributions from
first and second nearest neighbors as,
\begin{align}
\mathcal{A}_k^{NN}&=\mathcal{A}_k^{\lambda_2=0}= \frac{\lambda_1 (\lambda_1 - \mu \cos k)}{2(\lambda_1^2 + \mu^2 - 2\lambda_1 \mu \cos k)} \\
\mathcal{A}_k^{NNN}&=\mathcal{A}_k^{\lambda_1=0}= \frac{\lambda_2 (\lambda_2 - \mu \cos 2k)}{\lambda_2^2 + \mu^2 - 2\lambda_2 \mu \cos 2k} 
\end{align}
Using Eq.\ref{fracwind} we find the fractional contribution, i,e. $w_A=\frac{1}{2}$ is from
$\mathcal{A}_k^{\lambda_2=0}$. The integer contributions $w_B=1$ is from
$\mathcal{A}_k^{\lambda_1=0}$. The winding numbers $w_A$ and $w_B$ (loop A and B from left panel of Fig.\ref{dcir}) are
shown in right panel of Fig.\ref{dcir}. One should discard the fractional part and 
consider only the integer winding number, which correctly characterizes
the criticality. The results can be found in agreement with the method used
in the previous section. Similar argument can be applied to the 
transition between $w=0$ and $w=1$.
In this transition, since $\lambda_2=0$, from the right panel of Fig.\ref{dcir} one can observe a
separated integral forms taking the values $w_A=\frac{1}{2}$ and $w_B=0$. Here again 
the integer part correctly captures the trivial criticality with no localized MZMs.\\
According to the bulk-boundary correspondence, the topological invariant 
of the bulk which is generally an integer is equal to the 
number of MZMs present at the edge. Bulk-boundary correspondence does not provide 
any correspondence for fractional topological invariant number 
and the MZMs at the edge. The fractionality does not have any 
physical attribute at the edge i.e., there cannot be fractional MZMs.
Therefore it is 
necessary to consider only the integer part of the topological invariant and 
cast aside the fractional invariant number.
This way one can validate the bulk-boundary correspondence even at 
criticality. Characterization of critical phases by using both the methods, 
i.e calculation of topological invariant 
number by zeros of complex function associated with
the Hamiltonian and separation  of the 
integral forms to cast aside the fractional contribution
to the invariant number, 
are in agreement with each other. \\
In the further longer range coupling
the higher topological invariant number phases 
appear in the phase diagram. In addition to $w = 0,1,2$ regions,
$w = 3$ can also be found in appropriate regimes of the parameter space when the 
next-next-nearest neighbor interaction (NNNN) is included in 
the model Hamiltonian. Considering the phase transition from 
gapped phases $w=3$ to $w=2$ or from $w=3$ to $w=1$, two and 
one gapless MZMs appear at the criticality respectively.  
In general, when there is a topological phase transition occurring between 
$w=\alpha$ and $w=\beta$, number of MZMs that 
appear at the criticality are equal to the least 
topological invariant number among $\alpha$ and $\beta$ of 
the two gapped phases. 
\section{Summary and Discussions} \label{conclude}
\noindent In this work, we have attempted a study of MZMs in topological
superconducting Kitaev chain with longer range 
coupling.\@ We found 
fascinating behavior of MZMs which localize not only at gapped phase but 
also at the gap closing critical points. The MZMs at the criticality are
observed numerically and an alternate analytical method is used to characterize 
them. In spite of the invalidity of conventional topological invariant number at the
criticality, one can use a method of defining winding number using
number of zeros and poles of a complex function associated with model
Hamiltonian. This is successful in characterizing the triviality and
non-triviality of the criticality in the model. Clear agreement between
the winding number and the MZMs at criticality was found comparing the numerical
and analytical results. Moreover, by studying the system size
dependence, we found MZMs at criticality are as stable as that of MZMs at gapped phases. 
It is evident that the model posses topological
non-trivial criticalities which posses localized stable MZMs. 
These MZMs were found to appear 
only when either of the neighboring gapped phases are non-trivial and 
the number of MZMs is equal to the least number of invariant
number among the neighboring topological gapped phases.\\ 
A peculiar behavior was found on one of the critical line, 
which posses both topological trivial and non-trivial criticalities
separated by a multicritical point. MZMs are found to be 
localized at one regime on the critical line and delocalize into 
the bulk at other regime. This both topological and non-topological
nature on the single critical line causes the possibility of 
an unusual \textit{topological phase transition at quantum criticality},
which appear without bulk gap closing. The topological transition occurs
along the critical line and 
is independent of bulk gap
since at criticality the bulk gap is closed. 
This transition
occur through multicritical point, on the critical line, separating 
distinct criticalities.  Eigenvalue spectrum also shows the occurrence 
of topological phase transition on the critical line.\\
Integral definition of the invariant number defined by omitting
the gap closing point give rise to fractional winding numbers. 
To bring in the spirit of bulk-boundary correspondence
at the criticality, we have proposed a scheme to obtain
integer value of invariant number which efficiently characterize
the MZMs at criticality. We have shown explicitly that one can 
separate the fractional and integer contributions of Berry connection
to the invariant number. Since the fractional value does not 
account for any physical MZMs we discard it and retain only the 
integer contribution, which corresponds to the number of MZMs
at criticality. This validates the notion of bulk-boundary correspondence
even at criticality.
\section{acknowledgment}
	\noindent Authors would like to acknowledge Prof. Subir Sachdev, Prof. Brijesh Kumar and Dr. Abhishodh Prakash for the useful discussions.
	S.S. would like to acknowledge AMEF and DST (EMR/2017/000898) for the 
	support.\@ S.R., R.R.K., and Y.R.K. would like to acknowledge PPISR, 
	RRI library for the books and journals.\@ Authors would like to acknowledge Prof.\@ Sivaram and Prof. R. Srikanth for critically reviewing the manuscript. Finally authors would like to acknowledge ICTS for a
	useful discussion meetings and conferences.
\section{Appendix}
\subsection{A. Derivation of critical lines}
Energy dispersion relation of the Hamiltonian is given by,
\begin{equation}
E_k=\pm\sqrt{\chi_y(k)^2+\chi_z(k)^2},
\end{equation}
where $\chi_y(k)=2\lambda_1\sin k+2\lambda_2 \sin 2k$ and $\chi_z(k)=-2\lambda_1 \cos k-2\lambda_2\cos 2k+2\mu.$
Topological angle $\theta_k$ is given by,
\begin{equation}
\theta_k=\tan^{-1}\left(\frac{\chi_y(k)}{\chi_z(k)}\right).
\end{equation}
For critical line, energy spectrum should be zero i.e., $|E(k,\mu,\lambda_1,\lambda_2)|=0$, which occurs
at three distinct point points on the Brillouin zone. For all these points
non-zero component of the Hamiltonian ($\chi_{z}(k)$) gives the condition for the critical line.
For $k=0$, the analytical expression for the quantum critical line is $\lambda_2=\mu-\lambda_1$.
For the case of $k=\pm\pi$, we find the analytical expression of the another critical line as $\lambda_2=\mu+\lambda_1$.
For the case of $k=\cos^{-1}(-\frac{\lambda_1}{2\lambda_2})$, we find the analytical expression for the critical line $\lambda_2=-\mu$.
Thus three critical lines of the model Hamiltonian are $\lambda_2=\mu-\lambda_1$ and $\lambda_2=\mu+\lambda_1$ with $\lambda_2=-\mu$.
\subsection*{B. Complex analysis}
Model Hamiltonian written in the Majorana basis as,  
\begin{equation}
H=-i\left[ -\sum_{i=1}^{N} \mu b_ja_j+ \lambda_1 \sum_{i=1}^{N-1} b_ja_{j+1} + \lambda_2 \sum_{i=1}^{N-1} b_i a_{i+2}\right] .
\label{M} 
\end{equation}
where $a_j$ and $b_j$ are the Majorana operators with chemical potential $\mu$, nearest neighbor interaction $\lambda_1$ and next nearest neighbor interaction $\lambda_2$.
The general form of the Hamiltonian Eq.\ref{M} can be written as,
\begin{equation}
H = -i\sum_{\alpha=0,1,2} \left( \sum_{j=1}^{N-\alpha}\gamma_{\alpha} b_j a_{j+\alpha}\right) ,\label{com}
\end{equation}
The above Eq.\ref{com} can be brought to the complex form using Fourier transformation, $f(k)= \sum_{\alpha=0,1,2} \gamma_{\alpha} e^{ik\alpha}$, with $z = e^{ik}$. The complex function associated with the model Hamiltonian can be written as, 
\begin{equation} 
f(z) = \sum_{\alpha=0,1,2} \gamma_{\alpha} z^{\alpha}=-\mu + \lambda_1 z +\lambda_2 z^2,
\end{equation} 
where for $\gamma_{0,1,2}$ are respectively $-\mu$, $\lambda_1$ and $\lambda_2$. 
Here $f(z)$ has two solutions $z_1$ and $z_2$ which can be written as, 
\begin{equation}
z_{1,2}= \frac{-\lambda_1 \pm \sqrt{\lambda_1^2+4\mu \lambda_2}}{2\lambda_2}.
\label{s}
\end{equation}
These are two zeros of the function $f(z)$ (Eq.\ref{s}). Topological invariant ($w$) can be calculated using number of zeros of $f(z)$ inside the unit circle (since this model Hamiltonian does not have any poles). Zeros inside, on and outside the unit circle corresponds to topological, transition and non-topological phases respectively.
\subsection*{C. Edge mode analysis for both gapped and gapless phases}
The model Hamiltonian can be written as,
\begin{equation}
H_k = \chi_{z}(k) \sigma_z + \chi_{y}(k) \sigma_y,
\label{1}
\end{equation}
where $ \chi_{z} (k) = 2 \lambda_1 \cos k + 2 \lambda_2 \cos 2k - 2\mu,$ and $ \chi_{y} (k) = 2 \lambda_1 \sin k + 2 \lambda_2 \sin 2k.$\\
Substituting the exponential forms of $\cos k$ and $\sin k$, Eq.\ref{1} becomes, 
\begin{multline}
H = \left[ 2 \lambda_1 \frac{1}{2} (e^{-ik} + e^{ik}) + 2 \lambda_2 \frac{1}{2} (e^{-2ik} + e^{2ik} + 2 \mu)\right]\sigma_z\\ + i \left[ 2 \lambda_1 \frac{1}{2} (e^{ik} - e^{-ik}) + 2 \lambda_2 \frac{1}{2}(e^{2ik} - e^{-2ik}) \right] \sigma_y.
\label{2} 
\end{multline}  
We replace $e^{-ik} = e^{q}$, Eq.\ref{2} becomes, 
\begin{multline}
H = \left[ 2 \lambda_1 \frac{1}{2} (e^{q} + e^{-q}) + 2 \lambda_2 \frac{1}{2} (e^{2q} + e^{-2q} + 2 \mu)\right] \sigma_z\\ + i \left[ 2 \lambda_1 \frac{1}{2} (e^{-q} - e^{q}) + 2 \lambda_2 \frac{1}{2}(e^{-2q} - e^{2q}) \right] \sigma_y. 
\label{3} 
\end{multline}  
To find the zero mode solutions, we make $H^2 = 0$. 
By solving the Eq.\ref{3}, we get, 
\begin{equation}
2 \lambda_1 \cosh q + 2 \lambda_2 \cosh 2q - 2\mu = \pm  2 \lambda_1 \sinh q + 2 \lambda_2 \sinh 2q.
\label{4} 
\end{equation}
Eq.\ref{4} shows that there will be more than one solution. Considering the Eq.\ref{3} and squaring both sides with $H=0$, we get,  
\begin{multline}
\left(2 \lambda_1 \cosh q + 2 \lambda_2 \cosh 2q - 2\mu\right) \\+ i  \left(2 \lambda_1 \sinh q + 2 \lambda_2 \sinh 2q\right) = 0.
\label{5} 
\end{multline} 
Substituting back the exponential forms to the respective terms, we get, 
\begin{multline}
\left[ 2 \lambda_1 \frac{1}{2} (e^{q} + e^{-q}) + 2 \lambda_2 \frac{1}{2} (e^{2q} + e^{-2q} + 2 \mu)\right]\\ + \left[ 2 \lambda_1 \frac{1}{2} (e^{-q} - e^{q}) + 2 \lambda_2 \frac{1}{2}(e^{-2q} - e^{2q}) \right] = 0 
\label{6}
\end{multline}
Simplifying the Eq.\ref{6}, we end up with a quadratic equation,
\begin{multline}
2 \lambda_1 \frac{1}{2} e^{q}  + 2 \lambda_2 \frac{1}{2} e^{2q} + 2 \mu +  2 \lambda_1 \frac{1}{2} e^{q} + 2 \lambda_2 \frac{1}{2} e^{2q}  = 0.
\label{7}
\end{multline}
Simplifying the Eq.\ref{7} to a quadratic form,
\begin{equation} \lambda_2 e^{2q} +  \lambda_1 e^{q}  +  \mu = 0.
\label{8}
\end{equation}
The roots of this quadratic equation is given by, 
\begin{equation}
e^{q} = \frac{-\lambda_1 \pm \sqrt{\lambda_1^2 + 4 \lambda_2 \mu}}{2 \lambda_2} 
\label{9}
\end{equation}


\begin{thebibliography}{10}
	
	\bibitem{stanescu2016introduction}
	Tudor~D Stanescu. Introduction to Topological Quantum Matter \& Quantum
	Computation, CRC Press, 2016.
	
	\bibitem{wen2017colloquium}
	Xiao-Gang Wen.
	\ Reviews of Modern Physics, 89(2017) 041004.
	
	
	\bibitem{bansil2016colloquium}
	Arun Bansil, Hsin Lin, and Tanmoy Das.
	\ Reviews of Modern Physics, 88(2016) 021004.
	
	\bibitem{chen2019topological}
	Wei Chen and Manfred Sigrist.
	\ Advanced Topological Insulators, 239--280, 2019.
	
	\bibitem{landau}
	Lev Davidovich Landau.
	\ Ukr. J. Phys. 
	\ 11 (1937) 19-32.
	
	\bibitem{sachdev2007quantum}
	Subir Sachdev. 
	\ Handbook of Magnetism and Advanced Magnetic Materials. 2007.
	
	\bibitem{space}
	Robert-Jan Slager, Andrej Mesaros, Vladimir Juri{\v{c}}i{\'c}, and Jan Zaanen.
	\ Nat. Phys., 9(2013) 98.
	
	\bibitem{class}
	Ching-Kai Chiu, Jeffrey~CY Teo, Andreas~P Schnyder, and Shinsei Ryu.
	
	\ Rev. Mod. Phys., 88(2016) 035005.
	
	\bibitem{class1}
	Jorrit Kruthoff, Jan de~Boer, Jasper van Wezel, Charles~L Kane, and Robert-Jan
	Slager.
	\ Phys. Rev. X, 7(2017) 041069.
	
	\bibitem{hasan2010colloquium}
	M~Zahid Hasan and Charles~L Kane.
	\ Reviews of modern physics, 82(2010) 3045.
	
	\bibitem{pollmann2012symmetry}
	Frank Pollmann, Erez Berg, Ari~M Turner, and Masaki Oshikawa.
	\ Phys. Rev. B, 85(2012) 075125.
	
	\bibitem{altland1997nonstandard}
	Alexander Altland and Martin~R Zirnbauer.
	\ Phys. Rev. B, 55(1997) 1142.
	
	\bibitem{S2019}
	Rahul S, Ranjith~Kumar R, Y~R Kartik, Amitava Banerjee, and Sujit Sarkar.
	\ Physica Scripta, 94(2019) 115803.
	
	\bibitem{CHEN2020126168}
	Bo-Hung Chen and Dah-Wei Chiou.
	\ Physics Letters A, 384(2020) 126168.
	
	\bibitem{PhysRevE.100.020702}
	Mehmet Ramazanoglu, \ifmmode \mbox{\c{S}}\else~\c{S}\fi{}ener \"Oz\"onder, and
	Rumeysa Salc\ifmmode \imath \else~\i \fi{}.
	\ Phys. Rev. E, 100(2019) 020702.
	
	\bibitem{essin2011bulk}
	Andrew~M Essin and Victor Gurarie.
	\ Physical Review B, 84(2011) 125132.
	
	\bibitem{kitaev2001}
	A~Yu Kitaev.
	\ Phys. Usp., 44(2001) 131.
	
	\bibitem{su19795}
	WP~Su.
	\ 5. r. schrieffer and aj heeger.
	\ Phys. Rev. Lett., 42(1979) 1698.
	
	\bibitem{niu2012majorana}
	Yuezhen Niu, Suk~Bum Chung, Chen-Hsuan Hsu, Ipsita Mandal, S~Raghu, and Sudip
	Chakravarty.
	\ Phys. Rev. B, 85(2012) 035110.
	
	\bibitem{sarkar2018quantization}
	Sujit Sarkar.
	\ Sci. Rep., 8(2018) 5864.
	
	\bibitem{kopp2005criticality}
	Angela Kopp and Sudip Chakravarty.
	\ Nat. Phys., 1(2005) 53.
	
	\bibitem{kumar2020multi}
	Ranjith~R Kumar, YR~Kartik, S~Rahul, and Sujit Sarkar.
	\ Sci. Rep. 11(2021) 1-20.
	
	\bibitem{kartik2020topological}
	YR~Kartik, Ranjith~R Kumar, S~Rahul, Nilanjan Roy and Sujit Sarkar.
	\ arXiv:2009.04111, 2020.	
	
	\bibitem{massive}
	Oscar Viyuela, Davide Vodola, Guido Pupillo, and Miguel~Angel Martin-Delgado.
	\ Phys. Rev. B, 94(2016) 125121.
	
	\bibitem{lutchyn2018majorana}
	RM~Lutchyn, EPAM Bakkers, Leo~P Kouwenhoven, Peter Krogstrup, CM~Marcus, and
	Y~Oreg.
	\ Nat. Rev. Mater., 3(2018) 52--68.
	
	\bibitem{kane}
	Liang Fu and Charles~L Kane.
	\ Physical review letters, 100(2008) 096407.
	
	\bibitem{turner2013beyond}
	Ari~M Turner, Ashvin Vishwanath, and Chapter~Outline Head.
	\ Topological Insulators, 6 (2013) 293--324.
	
	\bibitem{wilczek2009majorana}
	Frank Wilczek.
	\ Nat. Phys., 5(2009) 614.
	
	\bibitem{nature2}
	Carlo Beenakker and Leo Kouwenhoven.
	\ Nat. Phys., 12(2016) 618.
	
	\bibitem{cheng2011majorana}
	Meng Cheng and Hong-Hao Tu.
	\ Phys. Rev. B, 84(2011) 094503.
	
	\bibitem{PhysRevB.92.075432}
	R.~S. Akzyanov, A.~L. Rakhmanov, A.~V. Rozhkov, and Franco Nori.
	\ Phys. Rev. B, 92(2015) 075432.	
	\bibitem{verresen2020topology}
	Ruben Verresen.
	\ arXiv preprint arXiv:2003.05453, 2020.
	
	\bibitem{jones2019asymptotic}
	Nick~G Jones and Ruben Verresen.
	\ J. Stat. Phys., pages 1--50, 2019.
	
	\bibitem{gapless}
	Anna Keselman and Erez Berg.
	\ Phys. Rev. 91(2015) 235309.
	
	\bibitem{kestner2011prediction}
	JP~Kestner, Bin Wang, Jay~D Sau, and S~Das Sarma.
	\ Phys. Rev. B, 83(2011) 174409.
	
	\bibitem{keselman2015gapless}
	Anna Keselman and Erez Berg.
	\ Phys. Rev. B, 91(2015) 235309.
	
	\bibitem{JIANG2018753}
	Hong-Chen Jiang, Zi-Xiang Li, Alexander Seidel, and Dung-Hai Lee.
	\ Science Bulletin, 63(2018) 753 -- 758.
	
	\bibitem{ganeshan2013topological}
	Sriram Ganeshan, Kai Sun, and S~Das Sarma.
	\ Physical review letters, 110(2013) 180403.
	
	\bibitem{cano2015chirality}
	Jennifer Cano, Meng Cheng, Maissam Barkeshli, David~J Clarke, and Chetan Nayak.
	\ Physical Review B, 92(2015) 195152.
	
	\bibitem{gao2019topological}
	Heng Gao, J{\"o}rn~WF Venderbos, Youngkuk Kim, and Andrew~M Rappe.
	\ Annual Review of Materials Research, 49(2019) 153--183.
	
	\bibitem{wang2019non}
	Huaiqiang Wang, Jiawei Ruan, and Haijun Zhang.
	\ Physical Review B, 99(2019) 075130.
	
	\bibitem{verresen2018topology}
	Ruben Verresen, Nick~G Jones, and Frank Pollmann.	
	\ Phys. Rev. Lett., 120(2018) 057001.
	
	\bibitem{anderson1958coherent}
	Philip~W Anderson.
	\ Physical review, 110(1958) 827.
	
	\bibitem{ablowitz2003complex}
	Ablowitz, M.J. and Fokas, A.S. \ Complex Variables: Introduction and Applications \ Cambridge University Press, 2003.
	
	
	\bibitem{10.2307/2397741}
	M.~V. Berry.
	\ Proceedings of the Royal Society of London. Series A,
	Mathematical and Physical Sciences, 392(1984) 45--57.
	
	
	
\end{thebibliography}
\end{document}